# Fast Complex Network Clustering Algorithm Using Agents

Di Jin
College of Computer Science and Technology
Jilin University
Changchun, China
jindi@email.jlu.edu.cn

Dayou Liu
College of Computer Science and Technology
Jilin University
Changchun, China
dyliu@jlu.edu.cn

Bo Yang
College of Computer Science and Technology
Jilin University
Changchun, China
ybo@jlu.edu.cn

Jie Liu
College of Computer Science and Technology
Jilin University
Changchun, China
liu_jie@jlu.edu.cn

*Abstract*—Recently, the sizes of networks are always very huge, and they take on distributed nature. Aiming at this kind of network clustering problem, in the sight of local view, this paper proposes a fast network clustering algorithm in which each node is regarded as an agent, and each agent tries to maximize its local function in order to optimize network modularity defined by function Q, rather than optimize function Q from the global view as traditional methods. Both the efficiency and effectiveness of this algorithm are tested against computer-generated and real-world networks. Experimental result shows that this algorithm not only has the ability of clustering large-scale networks, but also can attain very good clustering quality compared with the existing algorithms. Furthermore, the parameters of this algorithm are analyzed.

## I. INTRODUCTION

Many systems in real world exist in the form of network, such as social network, biological network, WWW network, etc., which are also called complex network. Complex network has been one of the most popular research areas in recent years due to its applicability to the wide scale of disciplines [1, 2, 3]. Getting abreast of the basic statistical properties of complex network such as "small world", "power law", etc., another property that has attracted particular attention is that of "community structure": the nodes in networks are often found to cluster into tightly-knit groups with a high density of within-group edges and a lower density of between-group edges [3]. The goal of network clustering algorithm is to uncover the real community structure in complex network.

The research on complex network clustering is of fundamental importance for both the theoretical significance and practical applications on analyzing network topology, comprehending network function, unfolding hidden law of network and forecasting network activities, which has been used in many areas, such as terrorist organization recognition, organization management, biological network analyzing, Web community mining, topic based Web document clustering, search engine, link prediction, etc [4].

So far, many network clustering algorithms have been developed. In terms of the basic strategies adopted by them, they are mainly fallen into two main categories: optimization and heuristic based methods. The former solves the network clustering problem by transforming it into an optimization problem and trying to find an optimal solution for a predefined objective function such as different kinds of cut criterions adopted by spectral methods [5, 6], and the network modularity employed in several algorithms [7, 8, 9]. On the contrary, there are no explicit optimization objectives in the heuristic based methods, and they solve the network clustering problem based on some intuitive assumptions or heuristic rules, such as Girvan-Newman (GN) algorithm [3], Clique Percolation Method (CPM) [10], Finding and Extracting Communities (FEC) [11], Label Propagation Algorithm (LPA) [12, 13], etc. Additionally, there are some other kinds of network clustering algorithms: local centralities based algorithm [14], agents based algorithm [15], locality mixing based algorithm [16], etc.

Nowadays, the sizes of networks are becoming larger and larger, and they take on distributed nature. Aiming at this problem, different from all existing approaches, this paper proposes a fast network clustering algorithm (FNCA) in the sight of local view, in which each node is regarded as an agent, and each agent tries to maximize its local function in order to optimize function Q, rather than optimize function Q from the global view as most methods. This algorithm treads each agent as a cluster at first; each agent updates its label according to the information offered by neighbors in order to maximize its local function at each iteration; algorithm stops when the labels of all agents in the network don't change.

## II. ALGORITHM

### A. The Main Idea

In 2004, aiming at complex network clustering problem, Newman proposed network modularity defined by function Q [17]. Function Q is the fraction of edges that fall within communities, minus the expected value of the same quantity if edges fall at random without regard for the community structure. One form is given by

$$Q = \frac{1}{2m} \sum_{ij} \left[ A_{ij} - \frac{k_i k_j}{2m} \right] \delta(c_i, c_j) \quad (1)$$

where $c_i$ is the community to which node $i$ is assigned, the $\delta$-function $\delta(u, v)$ is 1 if $u = v$ and 0 otherwise, $A =$

$(A_{ij})_{n \times n}$ is adjacency matrix of the network, $m = \frac{1}{2}\sum_{ij} A_{ij}$ is the number of edges in the network, $k_i$ is the degree of node $i$ [18].

In this paper, function Q is denoted by the sum of local function $f$ of all nodes in the network, and the computation of each node's function $f$ is only related to its own cluster. Consequently we transform (1) into (2), and give Theorem 1 based on (2).

$$Q = \frac{1}{2m}\sum_i f_i, \quad f_i = \sum_{j \in C_i}\left[A_{ij} - \frac{k_i k_j}{2m}\right] \quad (2)$$

**Theorem 1.** For $i = 1\ldots n$, function Q of complex network is monotone increasing following any $f_i$.

*Proof.* Let network $N$ and its current community structure $C$. Take any node $i$ in the network. Let the label of node $i$ be changed from $c_i$ to $c_i'$, and then the community structure of this network become $C'$, which makes $f$-value of each node whose label is $c_i$ or $c_i'$ be changed. Then the variation of Q-value can be computed as follows:
$Q(C') - Q(C) =$
$$\frac{1}{2m}\left(f_i(c_i') + (f_i(c_i') - (A_{ii} - \frac{k_i k_i}{2m})) - f_i(c_i) - (f_i(c_i) - (A_{ii} - \frac{k_i k_i}{2m}))\right)$$
then
$$Q(C') - Q(C) = \frac{1}{m}(f_i(c_i') - f_i(c_i)).$$

If $f_i(c_i') > f_i(c_i)$, there is $Q(C') > Q(C)$. Then this proof is finish.

Though function Q has resolution limit problem, recently it's still regarded as the measure of clustering quality in most researches on network clustering problem. This paper also takes it as objective function. As we can see from (2) and Theorem 1, function Q of complex network is monotone increasing following local funtion $f$ of any node. Starting from this theory, agents based fast network clustering algorithm—FNCA is proposed in this paper. In FNCA, each network node is regarded as an agent, and each agent has the ability of communicating with neighbors and updating its own label. At the beginning of FNCA, each agent is assigned to a unique label; during the execution of this algorithm, each agent attains local information by communicating with its neighbors, and tries to choose the label which can maximize its local function $f$ by using the information of its neighbors in order to optimize function Q; algorithm FNCA stops when all agents enter inactive state (which also means the labels of all agents don't change).

As we can see, FNCA tries to optimize function Q in the sight of local view rather than from the global view as most methods. In this algorithm, each node is regarded as an agent, and each agent tries to maximize its local function $f$ in order to optimize function Q. Therefore, FNCA is in essence a distributed gradient descent method, which makes use of the gradient of each node's function $f$. Additionally, because any agent in the network should have same label with one of its neighbors or it itself is a cluster, any agent can update its own label only by using the information of its neighbors, and needn't attain the information about global network community structure.

*B. Algorithm Description*

Based on the above section, a simple description of algorithm FNCA is as follows:
Procedure FNCA
begin
1    assign each node an agent in the network
2    assign each agent a unique label
3    do
4      for each $agent_i$ in the network
5      begin
6          $neighbors_i \leftarrow$ get neighbors of $agent_i$
7          $neighbors_i \leftarrow neighbors_i \cup agent_i$
8          $labels \leftarrow$ get unique labels from $neighbors_i$
9          find $label_j$ in $labels$ which can maximize function $f$ of $agent_i$
10        assign $label_j$ to $agent_i$
11   end
12 until labels of all agents don't change
end

At each iteration, each agent updates its label by maximizing its local function $f$. Though this strategy is very effective, it's also easy to fall into the trap of local optimal. In this paper, our resolution is to give a great probability $p$ as a parameter. Each agent chooses its best label under probability $p$, otherwise freely selects a better label than its current label. Additionally, this probability can be also set using simulated annealing algorithm.

If the labels of one agent's neighbors and the clusters related to them don't change at last iteration, then the label of this agent will not change at this iteration, but this condition is so strong. Additionally, we should have reason to believe that, if the labels of one agent's neighbors don't change, under a great probability the label of this agent will not change too. In order to speed up our algorithm, we make the agents which satisfy this weak condition enter sleeping (or inactive) state, which means they needn't update their labels any more. However, when some sleeping agent doesn't satisfy this condition, it should be awakened immediately, and begins to compute and update its label again. Experimental results show that, this operation can speed up our algorithm obviously, and doesn't affect its performance.

The termination condition of this algorithm is strong. However, generally speaking, the clustering solution is good enough before iteration number reaches 50 even though aiming at large-scale networks containing millions of nodes and edges. Therefore we can take the iteration number limitation as accessorial termination condition of our algorithm. Additionally, users can also give their acceptable Q-value as another accessorial termination condition.

As we can see, in this algorithm each agent has the ability of updating its label asynchronously and independently, and optimizing function Q to cluster networks only by using local information. Therefore we can extend this algorithm to complete distributed algorithm easily, and apply it to cluster the large-scale and dynamic networks in distributed situation.

*C. Time Complexity Analysis*

Let the iteration number limitation be $T$. In the network, let the number of total nodes be $n$, the number of total edges be $m$, the average degree of all nodes be $k$, and the average community size be $c$. We give some propositions as follows.

**Proposition 1.** The time complexity of algorithm FNCA can't be worse than $T*n*k*c$.

*Proof.* There are four loops in this algorithm. If we don't consider the speeding up strategy, the first three loops are obvious. As for the fourth loop, it shouldn't be greater than $c$ because the sizes of communities are always increasing in this algorithm, and the average community size is $c$ to the end. Furthermore, considering the speeding up strategy, this algorithm is more efficient.

**Proposition 2.** For large-scale networks, algorithm FNCA is near linear time.

*Proof.* We find out that the size of well community in large-scale networks is always a constant about 100 according to the analysis in literature [19], so $c$ can be considered as a constant. Because $T$ is a constant and $n*k$ is $m$, time complexity of this algorithm can be given by $O(m)$ for large-scale networks.

## III. EXPERIMENTS

In order to quantitatively analyze the performance of algorithm FNCA, we test this algorithm by using computer-generated and real-world networks, and analyze its parameters at last. All experiments are done on a single Dell Server (Intel(R) Xeon(R) CPU 5130 @ 2.00GHz 2.00GHz processor with 4Gbytes of main memory on Microsoft Windows Server 2003 OS). Our programming environment is Matlab 7.3.

*A. Computer-Generated Networks*

To test the performance of this method, we adopt random networks with known community structure, which have been used as benchmark problem for testing complex network clustering algorithms [3]. This kind of random network is defined as $RN$ ($C, s, d, z_{out}$), where $C$ is the number of communities, $s$ is the number of nodes in each community, $d$ is the degree of each nodes in the network, and each node has $z_{in}$ edges connecting it to members of its community and $z_{out}$ edges to members of other communities. As $z_{out}$ is increased from zero, community structures of networks become more diffused and the resulting networks pose greater and greater challenges to the community-finding algorithm. Especially, a network doesn't have community structure when $z_{out}$ is bigger than 8. The clustering solution of a random network is perfect only if each node is assigned to the correct community, and no communities are divided further. This measure is used to compute clustering accuracy in this paper.

In order to investigate the performance of FNCA, the accuracy of this algorithm is compared with GN algorithm [3], Fast Newman (FN) algorithm [7], CPM algorithm [10] and FEC algorithm [11]. Algorithm GN and FN are all classic network clustering algorithms which take function Q as objective function, and they are proposed by Newman. Algorithm CPM and FEC don't adopt any objective function, which were reported in Nature and TKDE respectively, and they are also very competitive algorithms at present. In our algorithm, parameter $p$ is set to be 0.95, iteration number limitation is set to be 100. Benchmark random network $RN$ (4, 32, 16, $z_{out}$) is used in this experiment. Fig. 1 (a) shows the experimental results. In Fig. 1 (a), y-axis denotes clustering accuracy, x-axis denotes $p_{in}$. For each $p_{in}$, for each algorithm, we compute the average accuracy through clustering 50 random networks. As we can see from this figure, our algorithm significantly outperforms the other four algorithms according to clustering accuracy. Moreover, as $z_{out}$ becomes larger and larger, the superiority of our algorithm becomes more and more significant. Especially, when $z_{out}$ equals 8, which means the number of within-community and between-community edges per vertex is the same, our algorithm can still correctly classify 97.97% of vertices into their correct communities, while the clustering accuracy of the other algorithms is low at this moment.

Computing speed is another very important criterion to evaluate the performance of clustering algorithms. Time complexity analysis of algorithm FNCA has been given in Sec. II C yet, so in this section we show the actual running time of our algorithm from experimental view in order to evaluate its efficiency. In our algorithm, parameter $p$ is set to be 0.95, iteration number limitation is set to be 50, and this parameters setting will be also used in all experiments in the following part of this paper. From literature [19] we find out that the size of well community in lager-scale networks is always about 100, so random network $RN$ ($C$, 100, 16, 5) containing $100C$ nodes and $800C$ edges is adopted in this experiment. Though community structure of the network is known, the number of communities can still be changed by $C$. Fig. 1 (b) shows the experimental results. In Fig. 1 (b), y-axis denotes the actual running time (second), x-axis denotes the scales of networks (number of nodes + number of edges). In this figure, the blue curve denotes running time of FNCA, and the red curve denotes running time of FNCA without speeding up strategy. As we can see from this result, our algorithm is near linear time, and the speeding up strategy is very useful.

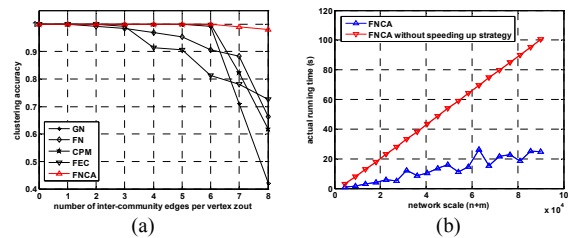

Figure 1. Testing the performance of FNCA against random networks. (a) Comparing FNCA with GN, FN, CPM and FEC in term of clustering accuracy; (b) The actual running time of FNCA on networks with different scales.

## B. Real-World Networks

We further test the performance of algorithm FNCA against many widely used real-world networks. There are not only small networks containing dozens of nodes, but also large-scale networks containing millions of nodes. A simple description of them is given by Table 1. As FNCA takes function Q as objective function, two algorithms which also employ Q as objective function are selected from Sec. III A. They are algorithm GN and FN. Comparing algorithm FNCA with these two algorithms against the networks described above, Table 2 shows the experimental results (empty cells correspond to computation time over 24 hours or out of memory). As we can see, our algorithm can attain good clustering solutions against large-scale networks containing millions of nodes and edges only within one hour. Its efficiency is much higher than GN and FN, and its clustering solution is good enough at the same time.

## C. Parameters Analysis

There are two parameters: probability $p$ and iteration number limitation $l$ in this algorithm. Taking the biggest four real-world networks WWW, amazon_2003_all, Web-google and Road-PA as example, this paper analyzes these two parameters now. For $l = 50$, we set the range of $p$ to be [0.9, 1] where its interval is 0.01, in order to test the sensitivity of parameter $p$ in our algorithm. Experimental results on the four networks are given by Fig. 2 (a). As we can see, aiming at different $p$, this algorithm can attain different clustering quality, but the variety is very small. Therefore, in general $p$ is set to be 0.95 in this paper.

For $p = 0.95$, $l$ is set to be 50. The trend that our algorithm's clustering quality varies with iteration number is given by Fig. 2 (b). As we can see, even against the networks containing millions of nodes and edges, our algorithm can attain good clustering solution within several iterations, but needs dozens of iterations in order to get a better result. Therefore, in general $l = 50$ is enough.

## IV. CONCLUSION

This paper tries to optimize objective function (function Q) of network clustering problem from local view, and proposes a fast network clustering algorithm by using agents. Experimental results show that this algorithm is valid. Our future work can be as follows. First, to improve the efficiency and effectiveness of this algorithm by proposing a better strategy to update the label of each agent; second, to extend this algorithm to complete distributed algorithm in order to cluster large-scale and dynamic networks in distributed situation.

ACKNOWLEDGMENT

This work was supported by National Natural Science Foundation of China under Grant Nos. 60503016, 60603030, 60873149, 60973088 the National High-Tech Research and Development Plan of China under Grant No. 2006AA10Z245, 2006AA10A309, and the Open Project Program of the National Laboratory of Pattern Recognition (NLPR). The second author Dayou Liu is the corresponding author.

REFERENCES

[1] D.J. Watts and S.H. Strogatz, "Collective Dynamics of Small-World Networks," Nature, vol. 393, June 1998, pp. 440-442, doi:10.1038/30918.
[2] A.L. Barabási, R. Albert, H. Jeong, and G. Bianconi, "Power-law distribution of the World Wide Web," Science, vol. 287, March 2000, pp. 2115, doi: 10.1126/science.287.5461.2115a.
[3] M. Girvan and M. E. J. Newman, "Community structure in social and biological networks," Proc. Natl. Acad. Sci. USA, vol. 99, Dec. 2002, pp. 7821-7826, doi: 10.1073/pnas.122653799.
[4] M. A. Porter, J. P. Onnela and P. J. Mucha, "Communities in Networks," Notices of the American Mathematical Society, vol. 56, Sep 2009, pp. 1082-1097, 1164-1166.
[5] M. Fiedler, "A Property of Eigenvectors of Nonnegative Symmetric Matrices and Its Application to Graph Theory," Czechoslovakian Mathematical Journal, vol. 25, 1975, pp. 619-637.
[6] J. Shi and J. Malik, "Normalized Cuts and Image Segmentation," IEEE Trans. on Pattern analysis and machine Intelligent, vol. 22, Aug. 2000, pp. 888-904, doi: 10.1109/34.868688.
[7] M. E. J. Newman and M. Girvan, "Finding and evaluating community structure in networks," Physical Review E, vol. 69, Feb. 2004, pp. 026113, doi: 10.1103/PhysRevE.69.026113.
[8] R. Guimera and L.A.N. Amaral, "Functional cartography of complex metabolic networks," Nature, vol. 433, Feb. 2005, pp. 895-900, doi:10.1038/nature03288.
[9] V.D. Blondel, J.L. Guillaume and R. Lambiotte, "Fast unfolding of communities in large networks," J STAT MECH-THEORY E, vol. 10, Oct. 2008, pp. 10008, doi: 10.1088/1742-5468/2008/10/P10008.
[10] G. Palla, I. Derényi, I. Farkas and T. Vicsek, "Uncovering the overlapping community structure of complex networks in nature and society," Nature, vol. 435, Jun. 2005, pp. 814-818, doi: 10.1038/nature03607.

TABLE I. REAL-WORLD NETWORKS USED IN OUR EXPERIMENTS.

| Networks | V(G) | E(G) | Description |
|---|---|---|---|
| karate | 34 | 78 | Zachary's karate club [20] |
| dolphin | 62 | 160 | Dolphin social network [21] |
| football | 115 | 613 | American College football[3] |
| world | 7,207 | 31,784 | Semantic network [10] |
| Cit-hep-th | 27,400 | 352,021 | Citations between physics (arxiv hep-ph) papers [22] |
| Protein homology | 30,727 | 1,206,654 | Similarity between two proteins using the BLAST algorithm [23] |
| arxiv | 56,276 | 315,921 | scientific collaboration networks [24] |
| Epinions | 75,877 | 405,739 | Who-trusts-whom network from epinions.com [25] |
| WWW | 325,729 | 1,090,108 | Edgeed WWW pages in the nd.edu domain [26] |
| amazon_2003_all | 473,315 | 3,505,519 | Amazon products from 2003 all [27] |
| Web-google | 855,802 | 4,291,352 | Web graph Google released in 2002 [28] |
| Road-PA | 1,087,562 | 1,541,514 | California road network [19] |

TABLE II. COMPARING FNCA (AVERAGE RESULTS OVER 50 RUNS) WITH GN AND FN.

| Q-value / time(s) | FN | GN | FNCA |
|---|---|---|---|
| karate | 0.2528/0.031 | 0.4013/0.1015 | 0.3991/0.0439 |
| dolphin | 0.3715/0.078 | 0.4706/0.3163 | 0.4854/0.0676 |
| football | 0.4549/0.125 | 0.5996/5.1464 | 0.6010/0.1035 |
| world | 0.3821/38.25 | - | 0.3813/18.2976 |
| Cit-hep-th | 0.5189/380.484 | - | 0.5866/135.4548 |
| Protein homology | 0.8612/771.703 | - | 0.8927/78.3524 |
| arxiv | 0.5953/2551.38 | - | 0.6122/100.7063 |
| Epinions | 0.3860/2036.98 | - | 0.3998/863.1282 |
| WWW | - | - | 0.7431/862.6942 |
| amazon_2003_all | - | - | 0.6552/2044.7 |
| Web-google | - | - | 0.6170/3225.1 |
| Road-PA | - | - | 0.5959/2326.1 |

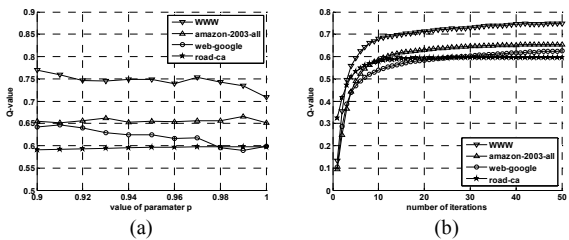

Figure 2. Parameters analysis. (a) Sensitivity analysis of parameter *p*; (b) Sensitivity analysis of parameter *l*.

[11] B. Yang, W. K. Cheung and J. Liu, "Community Mining from Signed Social Networks," IEEE Transactions on Knowledge and Data Engineering, vol. 19, Sept. 2007, pp. 1333-1348, doi: 10.1109/34.868688.

[12] U. N. Raghavan, R. Albert and S. Kumara, "Near linear time algorithm to detect community structures in large-scale networks," Physical Review E, vol. 76, Sept. 2007, pp. 036106, doi: 10.1103/PhysRevE.76.036106.

[13] I.X.Y. Leung, H. Pan, P. Li`o and J. Crowcroft, "Towards real time community detection in large networks," Phys. Rev. E, vol. 79, June 2009, pp. 066107, doi: 10.1103/PhysRevE.79.066107.

[14] B. Yang and J. Liu, "Discovering Global Network Communities Based on Local Centralities," ACM Trans. on the Web, vol. 2, Feb. 2008, pp. 1-32, doi: 10.1145/1326561.1326570.

[15] I. Gunes, and H. Bingol, "Community detection in complex networks using agents," 6th International Conference on Autonomous Agents and Multiagent Systems (AAMAS 07), ACM Press, May. 2007.

[16] B. Yang, J. Liu, J.F. Feng and D.Y. Liu, "On Modularity of Social Network Communities: The Spectral Characterization," Proceedings of 2008 IEEE/WIC/ACM joint Conferences on Web Intelligence and Intelligent Agent Technology (WI/IAT 08), IEEE Press, Dec. 2008, pp. 9-12, doi: 10.1109/WIIAT.2008.70.

[17] M. E. J. Newman and M. Girvan, "Finding and evaluating community structure in networks," Physical Review E, vol. 69, Feb. 2004, pp. 026113, doi: 10.1103/PhysRevE.69.026113.

[18] M.E.J. Newman, "Analysis of weighted networks," Physical. Review E, vol. 70, Nov. 2004, pp. 506131, doi: 10.1103/PhysRevE.70.056131.

[19] J. Leskovec, K.J. Lang, A. Dasgupta and M.W. Mahoney, "Community Structure in Large Networks: Natural Cluster Sizes and the Absence of Large Well-Defined Clusters," unpublished.

[20] W. W. Zachary, "An information flow model for conflict and fission in small groups," Journal of Anthropological Research, vol. 33, Apr. 1977, pp. 452-473.

[21] D. Lusseau, "The Emergent Properties of a Dolphin Social Network," Proceedings of the Royal Society of London Series B-Biological Sciences, Nov. 2003, pp. S186-S188, doi: 10.1098/rsbl.2003.0057.

[22] J. Gehrke, P. Ginsparg, and J. Kleinberg, "Overview of the 2003 KDD Cup," Proceedings of SIGKDD Explorations, ACM Press, Dec. 2003, pp. 149-151, doi:10.1145/980972.980992.

[23] University of Texas Bioinformatics, Proteomics, and Functional Genomics server, http://apropos.icmb.utexas.edu/lgl/.

[24] M.E.J. Newman, "The structure of scientific collaboration networks," Proc. Natl. Acad. Sci., vol. 98, Jan. 2001, pp. 404-409, doi: 10.1073/pnas.021544898.

[25] M. Richardson, R. Agrawal and P. Domingos, "Trust management for the semantic web," Proceedings of the 2nd International Semantic Web Conference, Springer Press, Jan. 2007, pp. 351-368, doi: 10.1007/b14287.

[26] Center for Complex Network Research, http://www.nd.edu/~networks/resources/.

[27] A. Clauset, M.E.J. Newman and C. Moore, "Finding community structure in very large networks," Physical Review E, vol. 70, Aug. 2004, pp. 066111, doi: 10.1103/PhysRevE.70.066111.

[28] Google Programming Contest, http://www.google.com/programming-contest/.